\documentstyle[11pt,newpasp,twoside,psfig]{article}
\begin{document}

\title{Dynamical Processes in the Central Kpc and Active Galactic Nuclei}
\author{Isaac Shlosman\altaffilmark{1}}
\affil{Racah Institute for Physics, Hebrew University,
Jerusalem 91904, Israel}

\altaffiltext{1}{Lady Davis Fellow, Hebrew University, Jerusalem, Israel. 
Permanent address: Department of
Physics \& Astronomy, University of Kentucky, Lexington, KY 40506-0055, USA} 

\setcounter{page}{1}
\index{Shlosman, I.}
\index{Shlosman, I.}

\begin{abstract}
We discuss different aspects of nested bar dynamics and its effect on the gas flow
and fueling of Active Galactic Nuclei. Specifically we focus on the dynamical
decoupling between the primary and secondary bars and the gas flow across the
bar-bar interface. We analyze the nuclear gaseous bar formation when gas gravity
can be neglected or when it dominates. Finally, we discuss the possible effect
of flat core, triaxial, dark halos on the formation of galactic bulges and 
supermassive black holes (SBHs)  and argue in favor of SBH-bulge-halo correlation.
\end{abstract}

\section{Introduction: Emerging Global Connection}
Departures from axial symmetry are destined to shorten the timescales
of secular and dynamical evolution in disk galaxies. Stellar bars, triaxial halos and 
tidal interactions play the important roles of driving such evolution on larger spatial 
scales, but their effect diminishes sharply within the central kpc. 
Is there 
any comparable non-axisymmetric morphology within the {\it circumnuclear} region which can 
impose gravitational torques, trigger bursts of star formation, and  
fuel the nonstellar activity of supermassive black holes (SBHs) --- ubiquitous, as 
recent observations confirm? What are the relevant processes which
maintain this morphology, and how much they affect
the galactic evolution?

Disk galaxies as a ``norm'' are barred or ovally distorted in the near-infrared
(NIR).
High-resolution ground-based instruments and the availability of the 
{\it HST\/} has allowed for the first time a meaningful analysis of central
morphology and kinematics. Although our knowledge of the inner regions of disk
galaxies is clearly incomplete, certain patterns in their dynamical evolution
and their relationship to larger and smaller spatial scales have emerged.

At least dynamically, the inner parts of disk galaxies can be defined
by the positions of the inner Lindblad resonances (ILRs), typically
at about 1~kpc from the center.
These resonances play an important role in filtering
density waves, propagating between the bar corotation
radius (CR) and the center. They are usually delineated by elevated star
formation rates and the concentration of molecular gas in nuclear rings,
which can serve as reservoirs for fueling the central activity.  

One can distinguish the disk and spheroidal components within the
central kpc. Surprisingly, the mass of the central SBH has been claimed to
correlate with the bulge properties
(e.g., reviews by Kormendy \& Gebhardt 2001; Merritt \& Ferrarese 2001), 
despite the common wisdom that black holes are fueled by disk accretion. 
However, formation of bulges can be tied directly to the
properties of triaxially-shaped halos, thus strongly hinting about the `global
connection' between smallest and largest spatial scales within the forming and 
evolving galaxy (Section~3 and El-Zant et al. 2002). 

Resolved plane morphology of central kpc in barred galaxies has revealed
so-far grand-design (mini) spirals and bars in addition to ``traditional'' disks
and bulges. 
We refer to large, kpc-scale
bars as ``primary,'' and to the sub-kpc bars as ``secondary.'' The
theoretical rationale behind these definitions is that secondary bars are
believed to form as a result of radial gas inflow along the large-scale bars
and, therefore, are expected to be confined within their ILRs
(Shlosman, Frank \& Begelman 1989). Below
we review morphologies of the circumnuclear region, analyze 
dynamical properties of nested bars,
their dynamical coupling to larger scales, and the resulting gas 
flows on scales of $\sim 10$~pc$-10$~kpc, as well as discuss the possible origin 
of SBH-bulge correlation within the context of triaxial halos.
 
\section{Nuclear Bars in Nested Bars: Gas, Stars and `Cocktails'}

In the powerful display of nonlinear dynamics, many galaxies exhibit double
stellar bars, which can tumble with different or identical pattern speeds. 
Although examples have been known for nearly three decades, the dynamical
importance of nested bars has been first pointed out much later.
The largest sample of (112) disk galaxies analyzed so far
for this purpose reveals a substantial fraction of nested bars,
probably in excess of $20-25$\%. Even more interestingly, about 1/3 of {\it barred} 
galaxies host a second bar (Laine et al. 2002).
The gas contents of nuclear bars vary. 
In some cases, the cold gas can be dynamically important, as evident from the 
interferometric 2.6~mm CO emission and NIR lines of H$_2$.
Depending on the gas fraction contributing to the gravitational
potential, one can distinguish  stellar-, gas-dominated (i.e., gaseous) and 
mixed nuclear bars. The nuclear {\it gaseous} bars have no large-scale counterparts.

Nuclear bars extend the action of gravitational torques to smaller spatial scales.
Their relevance for the AGN depends of course on the availability
of the `fuel' (i.e., gas) which loses angular momentum and falls toward the center.  
The gas-dominated nuclear bars can be especially important for this process, and therefore
we discuss them in some detail below.
Probably the most intriguing property of nested bars is their theoretically
anticipated stage of a dynamical decoupling, when each bar exhibits a
different pattern speed (Shlosman et al. 1989). Several aspects
of this problem are analyzed below. 

\noindent\underline{\it Gas Flow in Nested Bars.}\ \ 
Here we assume that nested bars tumble with different pattern speeds, 
$\Omega_s > \Omega_p$, where `p' stands for primary and `s' 
--- for secondary (sub-kpc) bar.
Arguments dealing with chaos minimization in such time-dependent system will
lead to certain limits, specifically that CR of the secondary bar 
must lie in the vicinity of the primary bar ILR, constraining $\Omega_s$
(e.g., Pfenniger \& Norman 1990). Such dynamical configuration, in
principle, poses a problem for uninterrupting gas inflow towards smaller
radii. According to this argument, the gas flow is repelled at the bar CR, 
inwards or
outwards, because of the rim formed by the effective potential there, and
hence may not cross the ILR/CR (i.e., bar-bar interface).
Shlosman et al. (1989) have argued, in essence, that it is the gas
self-gravity that overcomes repulsion by modifying the underlying
potential. But even in the limit of neglibigle gravity in the gas,
the flow is capable of crossing the bar-bar interface, although not in a steady
manner and only for a restricted range of azimuthal angles (Shlosman \& Heller
2002).  

The pattern of shock dissipation in nested bars can be inferred from Fig.~1.
It allows one to separate the incoming large-scale shocks from those driven
by the secondary bar. Note that two systems of spiral shocks occur,
each associated with the corresponding bar. 
The interaction between these shock systems shows detachment when the bars are
perpendicular and attachment when they are aligned with each other. The shapes
of the shocks depend on the angle between the bars. 

Several factors characterize the gas dynamics in the {\it decoupled}
bars: $(i)$ the time-dependent nature of the potential;
$(ii)$ the nonsteady gas injection into the secondary bar which proceeds
through the primary shocks penetrating the bar-bar interface.
This phenomenon is absent at the CR of the primary bars. Unstable
orbits in the interface region preclude the secondary bars from extending to
their CR (El-Zant \& Shlosman 2002, in preparation); $(iii)$ a fast-tumbling secondary bar
which prevents the secondary ILRs from forming. Even
in the case of a long-lived decoupled phase secondary bars are not expected
to slow down. In fact, the gas inflow across the interface and the resulting central
concentration can speed-up the bar (Heller, Noguchi, \&
Shlosman 1993, unpublished). The low-Mach-number gas flow is well organized
and capable of following orbits within the bar with little dissipation. Non-linear
orbit analysis reveals that, in the deep interior of the secondary bar, the
$x_1$ orbits have a mild ellipticity and no end-loops. This result is robust.
No offset large-scale shocks form under these conditions.

\begin{figure}[ht!!!!!!!!!!!!!!!!!!!!!!!!!]
\vbox to4.4in{\rule{0pt}{4.4in}}
\includegraphics{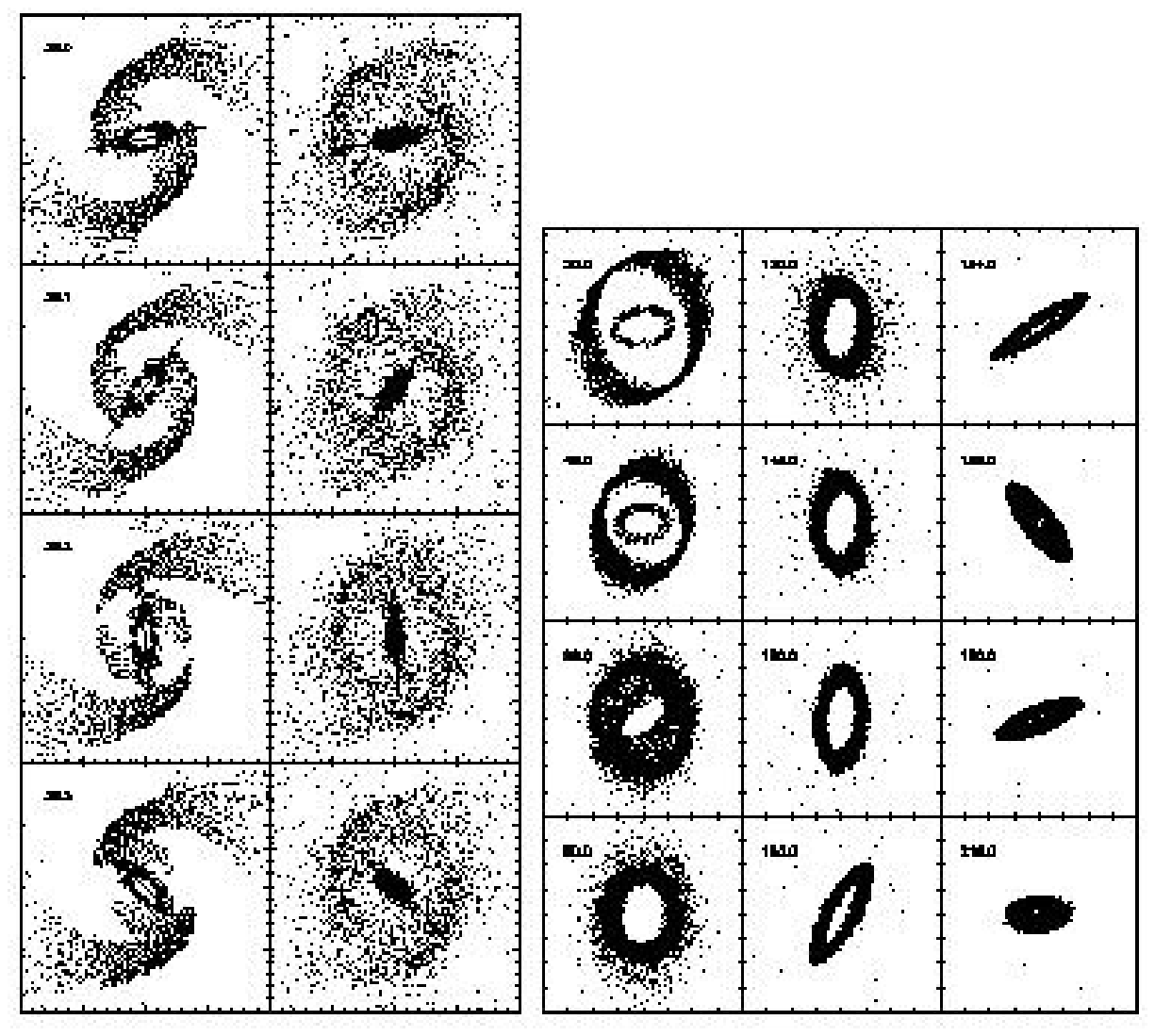} 
\parindent 0.8truecm\parbox{4.7in}{\underline{{\it Left:\/} Figure 1.} Pattern
of shock dissipation (left) and density evolution (right) in the central kpc,
in the frame of reference of the primary bar (horizontal). Positions of the
secondary bar and its length are indicated by a straight line. All rotation is
counter-clockwise. Particles in the left column have greater than average
dissipation rate. Note the sharply reduced dissipation in the
innermost secondary bar and ``limb brightening'' enveloping it. Also visible
are two dissipative systems --- the shocks in primary and
secondary bars (Shlosman \& Heller 2002).
\underline{{\it Right:\/} Figure 2.} Time evolution of the low-viscosity model:
2D SPH simulation in the background potential of a barred disk
galaxy (face on). The gas response to the bar torquing is displayed in
the primary-bar (horizontal) frame. The gas rotation is
counter-clockwise. Note a fast evolution after $t\sim 150$, when the secondary
bar decouples and swings clockwise! The bar is ``captured'' again at $t\sim
211$. Time is given in units of dynamical time. This animation sequence
and others are available in the online edition of Heller et al. (2001).} 
\end{figure}     

Knapen et al. (1995) have analyzed the shock  dissipation in a self-consistent
potential of `live' stars and gas {\it before} the onset of
decoupling, when both bars tumble with the same pattern speeds, and when the
gas gravity is accounted for. No offset shocks have been found in this
configuration either. These results indicate that gas inflows stagnate within
the inner parts of fast rotating nuclear bars, forming nuclear disks 
{\it if gas gravity is neglected} 
--- a condition similar to having a low surface density gas. 

We conclude that compelling arguments show that no large-scale shocks and 
consequently no offset dust lanes
will form inside secondary nuclear bars either when they are dynamically coupled
and spin with the same pattern speeds as the primary bars, or dynamically decoupled,
spinning much faster, if gas gravity is neglected. 
However, the fate of the gas settling inside the
nuclear bars cannot be decided without invoking global gravitational
effects in the gas which will completely change the nature of the flow.

\noindent\underline{\it Dynamical Decoupling of Nested Bars.}\ \
Decoupling in nested bars is indirectly supported by the observed random 
orientation of primary and secondary bars.
To complicate the matter,
both bars can corotate, being completely synchronized.
This configuration of nearly orthogonal bars may be a precursor to the future
decoupled phase.  
The gas responding to the
gravitational torques from the primary bar flows towards the center and
encounters the $x_2$ orbits, which it 
populates. The forming
secondary bar may be further strengthened by the gas gravity,
or the amount of gas accumulating in the ILR
resonance region may be insufficient to cause this runaway. 
The computational effort has so far gone into analyzing
self-gravitating systems (e.g., Friedli 1999; Shlosman 1999).

If the secondary bar forms via gravitational instability (in
stellar or gaseous disks), it will spin in the direction of the primary bar
with $\Omega_s>\Omega_p$ (Shlosman et al. 1989; Friedli \& Martinet 1993;
Heller \& Shlosman 1994). The  presence of gas appears to be
imperative for this to occur. Both bars are  dynamically {\sl decoupled} and
the angle between them becomes arbitrary. The secondary bar can also rotate in
the opposite sense to the primary bar, resulting from merging
--- a non-recurrent configuration. 

\noindent\underline{\it Decoupling of Non-Self-Gravitating Bars.}\ \ 
The actual
degree of viscosity in the ISM is largely unknown. Heller, Shlosman \& 
Englmaier (2001) have investigated the effect of viscosity on the gas settling 
in the ILR region of a single large-scale stellar bar. 
The most spectacular evolution occurred in the low-viscosity model, although
all the models showed similar initial evolution, during which the gas
accumulated in a double ring, corresponding roughly to two
ILRs. In all the models the rings interacted hydrodynamically and merged (Fig.~2), 
forming a single oval-shaped ring corotating with the primary bar and
leading it by $\phi_{\rm dec}$ --- depending on
the gas viscosity. The remaining ring becomes increasingly
oval and barlike (Fig.~2), its pattern speed changes abruptly, and it
swings  towards the primary bar, {\sl against} the direction of rotation. In 
the inertial frame, this forming gaseous bar spins in
the same sense as the main bar, albeit with $\Omega_s<\Omega_p$, but in the 
primary-bar frame it tumbles retrograde (!)
for about 60 dynamical times, $\sim$ 2--3 $\times
10^9$~yr, until it is captured again by the primary. The shape of
the decoupled bar and $\Omega_s$ depend on bar orientation. The eccentricity,
$\epsilon$, reaches a maximum when bars are aligned. In standard and
high-viscosity models, the secondary bar only librates about the primary.

\stepcounter{figure}
\stepcounter{figure}
\begin{figure}[ht!!!!!!!!!!!!!!!]
\vbox to3.0in{\rule{0pt}{3.0in}}
\includegraphics{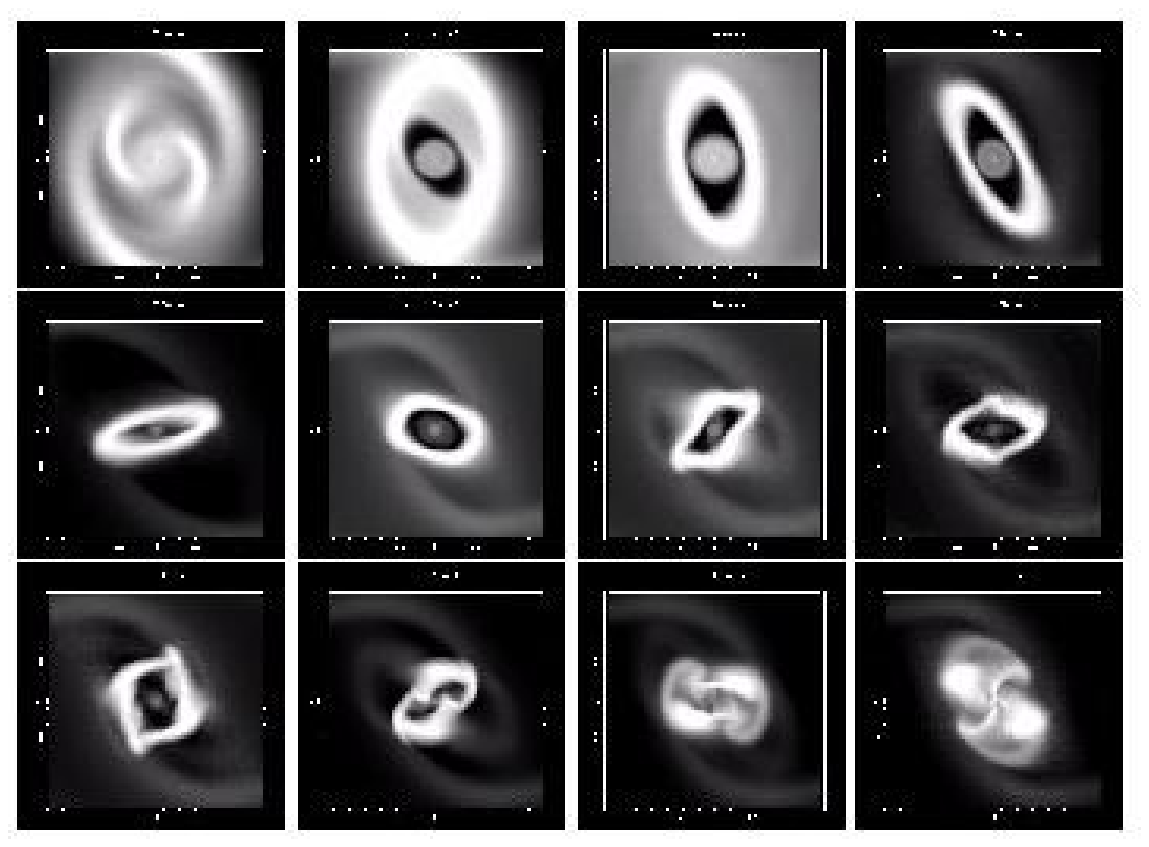}
\caption{Formation and decoupling of a self-gravitating gaseous bar (central kpc) 
in the reference frame of the large-scale stellar bar (horizontal). All rotation 
is clockwise. The time increases to the right. The first three frames: formation
of elongated ring perpendicular to the stellar bar and at its inner ILR; next 
three frames: `retrograde,' $\Omega_s<\Omega_p$  {\it non}-self-gravitational 
decoupling (compare with Fig.~2);
last six frames: prograde $\Omega_s>\Omega_p$ gravitational modes $m=2, 4$ develop 
and take over. This last stage is accompanied by a `catastrophic' growth of the 
central seed BH 
(Englmaier \& Shlosman 2001, unpublished).
}
\end{figure}

The key to understanding this behavior lies in the distribution of gas
particles with Jacobi energy, $E_{\rm J}$. After
merging the ring is positioned close to the energy where the transition from
$x_2$ to $x_1$ (at the inner ILR) occurs. The exact value of this transition
energy is model-dependent, but this is of no importance to the
essence of the decoupling. The crucial difference between the
models comes from $(i)$ the position of the forming gaseous bar on the $E_{\rm
J}$ axis after ring merging and $(ii)$ the value of $\phi_{\rm dec}$.

As the gaseous bar forms at $\phi_{\rm dec}$ to the primary, 
gravitational torques act to align the bars. In the low-viscosity model, 
gaseous bar resides on purely $x_2$ orbits and, therefore, responds to the
torque by speeding up its precession while being pulled backwards,
until it is almost at right angles to the bar potential valley. The
decoupling happens abruptly when $\sim1/2$ of the gas finds itself at $E_{\rm
J}$ below the inner ILR. The absence of $x_2$ orbits at these $E_{\rm J}$
means that the gas loses its stable orientation along the
primary-bar minor axis. The gaseous bar has a much smaller $\epsilon$ in the
fourth quadrant. Such an asymmetry with respect
to the primary ensures that the torques from
the primary bar are smaller in the fourth quadrant. But only for the least
viscous model this becomes crucial, and the torques are
unable to confine the bar oscillation, which continues for a full swing of
$2\pi$. The nuclear bar is trapped again, after few rotations.

Correlation between $\epsilon$ of the gaseous bar and
its orientation can be tested observationally. Two additional effects should have
observational consequence.
First, the gas will cross the inner ILR
on a dynamical timescale. Shlosman et al. (1989) pointed out that  ILRs
present a problem for radial gas inflow because the gas can stagnate there. A
solution was suggested in the form of a global gravitational instability
in the nuclear ring or disk, which will generate gravitational torques in the
gas, driving it further in. Recently  Sellwood \& Moore (1999) 
resurrected the idea that ILRs would ``choke'' the gas inflow. Fig.~2 shows that 
even non-self-gravitating nuclear rings are prone to dynamical
instability which drives the gas inward. Accounting for the gas gravity dramatically
affects the stability of gas at the inner ILR, as discussed below (see also Fig.~3). 
This inflow
is expected to be accompanied by star formation along the
molecular bar, having a quasi-periodic, bursting character.      

\noindent\underline{\it Decoupling of Self-Gravitating Bars.}\ \
Gas gravity
plays a crucial role in the formation and decoupling of 
bars, securing $\Omega_{\rm s} > \Omega_{\rm p}$.
Simulations which tackle  this issue
have been very limited so far. Friedli \& Martinet (1993, and priv.
communication) and Combes (1994) experienced difficulties in decoupling pure
stellar or gaseous bars. The inner bar in this case is very transient, which
may be a result of an insuffient number of particles.  For a mixed
system the decoupled stage is more prolonged. Simulations explicitly
demonstrate the necessity for gas to be present, and
confirm that an increased central-mass concentration is important for the
dynamical separation of the outer and inner parts. This can be achieved by
moving the gas along the large-scale bar, accumulating it inside the ILR on
the $x_2$ orbits, modifying the local potential and forming a double ILR. The
gravity of the gas settling onto these orbits is sufficient to ``drag'' the
stars along, but such a  configuration still corotates with the primary.
Settling onto the $x_2$ orbits depends upon gas sound
speed and viscosity. When the gas is too viscous or hot, it will avoid the
ILRs completely and remain on the $x_1$ orbits aligned with the bar.
Moderately viscous gas will settle onto the innermost $x_2$ orbits. The
evolution of nuclear regions in disks, therefore, depends on the (unknown)
equation of state of the ISM. The inclusion of star formation in nuclear rings
has already demonstrated how the resulting increase in viscosity leads to the
mass transfer across the ILRs (Knapen et al. 1995).

Shlosman (2001) have described simulations which have accounted for
gravitational effects in the gas. Ring evolution was similar to that
without gas gravity, including the swing towards the primary bar. But this
was followed by the rapid growth of {\sl prograde,} $\Omega_s>\Omega_p$, 
self-gravitating modes with
$m=2$ and 4 with the pattern speed much larger than $\Omega_{\rm p}$, resulting
in an avalanche-type inflow (Fig.~3). A model with the seed 
SBH revealed that inflowing gas feeds the SBH at
peak rates, increasing its mass tenfold. This model shows that, at least theoretically,
{\it gaseous bars can dump the fuel at the very vicinity of the SBH.}

\section{Central Black Holes and Bulge-Halo Connection}

Recent observations of galactic centers show clear correlations between the masses
of the SBHs, $M_\bullet$, and masses and dispersion velocities of surrounding bulges,
$M_B$ and $\sigma_B$, respectively.  
So far this is the most conclusive evidence that the very centers of galaxies are
coupled in their evolution with the main bodies of their hosts. Analysis of the
available data has led to the conclusion that the {\rm log}~$M_\bullet-{\rm~log}~\sigma_B$
relation has a slope of 4 (Tremaine et al. 2002). El-Zant et al. (2002) have attempted to
provide a physical explanation to this tantalizing phenomenon, based on the 
interaction between dark halos and baryonic components settling in their midst.  
In this formulation, the halo properties determine those of the bulge and the SBH. 

\begin{figure}[ht!!!!!!!!!!!!!!!]
\vbox to2.in{\rule{0pt}{2.in}}
\includegraphics{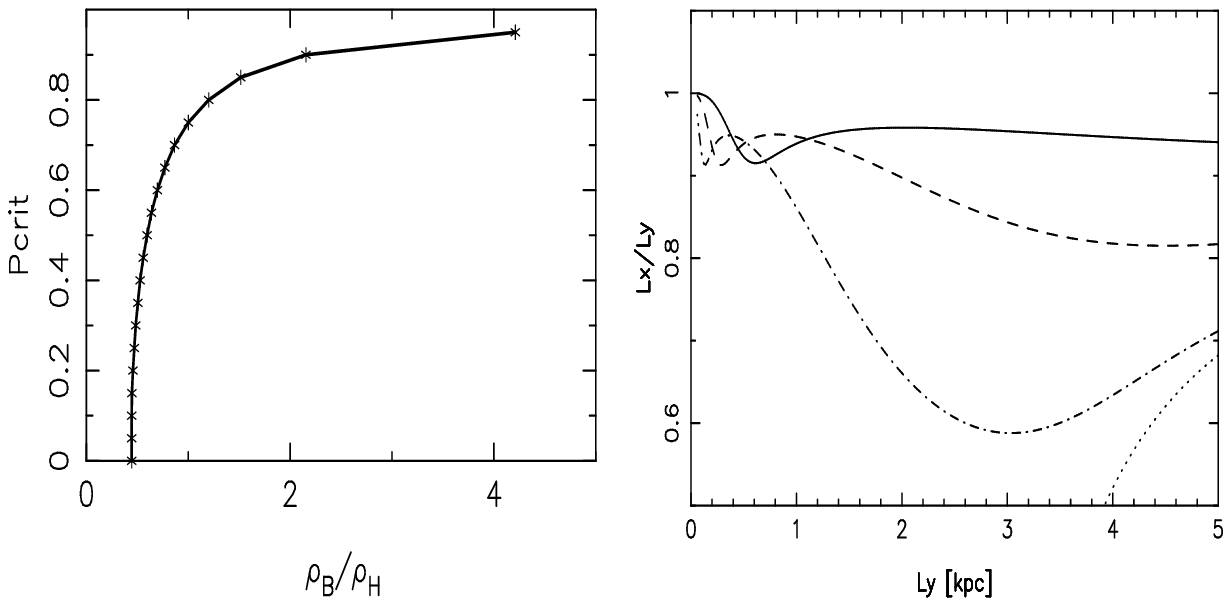}
\caption{\underline{\it (Left).}\/
Variation of the minimal critical axis ratio $p_{\rm crit}$ of the loop orbits as a function
of relative bulge density $\rho_B/\rho_H$.
\underline{Figure~5 {\it (Right).}\/} Axis ratios of closed loop orbits
$p=L_x/L_y$ as a function of $L_y$. The dotted line corresponds to a system with 
only a halo contribution. The solid line refers to a system with a Plummer ``bulge'' 
with $M_B/M_H=1$ and core radius $R_B=2$~kpc; the dashed line --- $M_B/M_H=0.1$
and $R_B=0.93$~kpc; the dashed-dotted line --- $M_B/M_H=0.01$ and $R_B=0.43$~kpc. 
The bulge core density is $\rho_B/\rho_H\sim 5.25$ and $K=10^{-3}$ (El-Zant et al.
2002).}
\end{figure}

The basic assumption is that of flat core triaxial halos subject to baryon infall
(e.g., El-Zant, Shlosman \& Hoffman 2001).
The flat cores do {\it not} support loop orbits and 
stable gas motions. It is the baryon influx which perturbs the background harmonic
core and leads to the appearance of progressively less oval loops. At the same
time, the contracting baryonic component becomes self-gravitating. Interestingly,
these two phenomena correlate (Fig.~4), so one expects that the onset of massive star
formation coincides with termination of the dynamical infall, forming a bulge-like
configuration. Unless the bulge is cuspy, the loop orbits are still absent deep
inside, where the infall proceeds until the formation of a central 
SBH. 

This settling of a baryonic matter in the mildly triaxial halos can in principle lead
to the simultaneous formation of the SBH-bulge system with a characteristic bulge
density, set by the density in the background halo. While the formation of round
loop orbits in the vicinity of the SBH will choke the dynamical infall, the bulge
will still grow, until a similar process will terminate this. The
inner `bottleneck' depends on the bulge-to-halo density ratio, $\rho_B/\rho_H$, while the outer
bottleneck is governed by the bulge-to-halo mass ratio --- limiting the SBH-to-bulge
mass ratio $K$ to largely between $10^{-4}-10^{-2}$ (Fig.~5). Such a large spread in $K$
is indeed observed. The spread in $M_\bullet-\sigma_B$ relation can be minimized
if the halo follows the Faber-Jackson scaling (Faber \& Jackson 1976). 

The most important prediction from modeling this nonlinear process is that the bulge 
and SBH parameters are determined by the halo properties. Within this framework, 
larger and more massive halo cores are expected to produce, on average, larger and more
massive bulges. Disks will form outside the halo core, or later, when
the baryons destroy the core. If the core is very
large, most of the baryonic material is consumed in the first phase and no
significant  disk forms. This effect is expected to be prominent in larger
mass cores, since if these follow the Faber-Jackson relation, more massive
halos should have  proportionally larger cores. Other predictions include 
the requirement that the  average density of the bulge in the central region
should be close to  that of the halo core. There is also a minimal bulge mass
associated with a given core,  although this varies significantly with
the critical loop eccentricity assumed. A number of additional  consequences
for galaxy formation and evolution will be discussed elsewhere.

\section{Conclusions} 

The subjects which have been emphasized in this review are 
nested bars and the SBH-bulge correlation. 
The nested bars are interesting for two main reasons, the possiblity to find them
in the dynamically decoupled stage, and being a prime channel to fuel the AGN.
Clearly, the ability of nested bars to trigger the gas inflow toward the very
center depends on the gas fraction within the central kpc. Gaseous bars can dump
the fuel at the very vicinity of the SBH. Neglecting the gravitational
effects in this gas artificially terminates its subsequent evolution.

The origin of the SBH-bulge correlations can be the manisfestation
of a global connection between the dark halos and the baryon-dominated
luminous parts of galaxies. The flat-core triaxial halos provide the possible
physical mechanism
which can shape these inner parts, depending on the properties of dark matter
component, in particular whether the latter exhibits the Faber-Jackson scaling. 
The halo shapes will have also a profound effect on the formation and subsequent
growth of disk component and its stability to bar formation at higher redshifts.

\acknowledgements
I am grateful to my collaborators on the subjects discussed here, 
and to numerous colleagues for exciting discussions.
Partially supported by NAG 5-10823, HST-AR-09546.01-A and AST-0206251 grants.

\end{document}